\documentclass[aps, prd, twocolumn, a4paper]{revtex4-1}
\usepackage{hyperref}
\usepackage{graphicx}
\usepackage{amsmath}
\usepackage{xcolor}
\usepackage{float}
\usepackage{amsmath}
\newcommand{\be}{\begin{equation}}
\newcommand{\ee}{\end{equation}}
\newcommand{\ba}{\begin{eqnarray}}
\newcommand{\ea}{\end{eqnarray}}

\begin{document}
\title{Quarkonium dissociation properties of hot QCD medium at momentum-anisotropy in the N-dimensional space using Quasi-particle Debye mass with 
finite baryonic chemical potential}
 \author{Vineet Kumar Agotiya$^{a}$}
\email{agotiya81@gmail.com}
 \author{Siddhartha Solanki$^{a}$}
\email{siddharthasolanki2020@gmail.com}
\author{Manohar Lal$^{a}$} 
\email{manoharlalphd2019@gmail.com}
\affiliation{$^a$Department of Physics, Central University of Jharkhand
 Ranchi, India, 835 222}
\begin{abstract}

The analytical exact iteration method (AEIM) have been used to calculate the $N$-dimensional radial schrodinger equation with
the real part of complex-valued potential and it is generalized to the finite value of anisotropy (\mbox{\boldmath$\xi$}), temperature 
and baryonic chemical potential. In $N$-dimensional space the energy eigen values have been calculated for any states (n,l).
The present results have been used to study the properties of quarkonium states (i.e, the binding energy and mass spectra
in the $N$-dimensional space). 
The influences of anisotropy (\mbox{\boldmath$\xi$}) on the dissociation temperature($T_D$) has been also calculated for the ground state of quarkonia at N=$0$ and $\mu=300MeV$. 
The anisotropy in oblate case, thus obtained, makes the dissociation temperature($T_D$) lower as compared to the isotropic case . We also seen that Leading order Debye mass is greater than that of Quasi-particle Debye mass after introducing the baryonic chemical potential.\\
{\bf PACS}:~~ 25.75.-q; 24.85.+p; 12.38.Mh
\\
{\bf Keywords} :  Schrodinger equation, real part of complex-valued potential, Debye mass, Momentum anisotropy,
Heavy quarkonia, dimensionality number and baryonic chemical potential.
\end{abstract}
\maketitle

\section{Introduction}

The study of the system of heavy quarkonia (such as Bottomonium and Charmonium) have played an important role for the standard model~\cite{S.M.Kuchin}, it also useful for the quantifiable test of Quantum Chromo-Dynamics (QCD), hence, is a topic of resurgent interest for both the theoretical and experimental high energy physicists.
The radial Schrodinger equation has been~\cite{R.Kumar} solved with the real part of complex valued potential, and the solution can be used to understand many phenomena 
in the study of atomic and molecular physics, spectroscopy (Hadronic as well as Molecular), nuclear physics and also in high energy physics which are not yet understood.
Most of the studies on the study of $N$-dimensional space problem ~\cite{TDas}, have been focused on the lower spatial 
or dimensional space~\cite{AAlJamel}.
The consequences of $N$-dimensional space have been considered on the energy levels of the system of quantum mechanics~\cite{ANIkot}. In the $N$-dimensional space,
the study of the harmonic oscillator~\cite{SMAlJaber} and hydrogen atom~\cite{AlJaber} have also been done.
Additionally, in $N$-dimensional space, the Schrodinger equation has also been solved for potentials such as Cornell potential~\cite{MAbushady}, fourth order 
inverse power potential~\cite{GRKhan}, Mie type potential~\cite{SIkhdair}, Kratzer potential~\cite{SMIkhdair}, Coulomb potential~\cite{GChen}, Energy 
dependent potential~\cite{HHassanabadi}, Global potential~\cite{GRBoroun}, Hua potential~\cite{HHassanabadi} and Harmonic potential~\cite{TDas} etc.
In these studies, the solutions of $N$-dimensional Schrodinger equation for fixed chemical baryonic potential at finite values of temperature
have been used to define the properties of Charmonium and Bottomonium mesons.
The radial schrodinger equation has been solved by using numerous methods such as, power
series method~\cite{S.M.Kuchin}, Hill determinant
method~\cite{RNChoudhury}, numerical methods~\cite{LGrIxaru, PSandin, ZWang}, quasi-linearization method (QLM)~\cite{EZLiverts}, point canonical transformation (PCT)~\cite{RDe} and operator algebric method~\cite{JJSakurai} etc. But the N-dimensional 
radial Schrodinger equation has also ben solved by other methods like, Nikiforov-Uvarov (UV) method~\cite{ANIkot, DAgboola, HHassanabadi}, 
power series techniques~\cite{RKumar12}, Laplace transformation methods~\cite{TDas, GChen}, Asymptotic iteration method (AIM)~\cite{R.Kumar}, SUSQM method, 
Analytical exact iteration method (AEIM)~\cite{EMKhokha} and Pekeris type approximation~\cite{HHassanabadi, HRahimov} etc.
The ground state of quarkonium dissociation rates have been studied by direct continuum of thermal activation, tunnelling, binding energy and 
phase shift scattering in the hot quark-gluon plasma (QGP) system for the eigen states of lowest values~\cite{PSandin, DKharzeev}. 
Vija and Thoma~\cite{HVija} extended perturbated gauge theory for the collisional energy loss in QGP at finite chemical potential and temperature.
The Bottomonium and Charmonium dissociation has been studied through the correction of cornell potential via hard thermal loop resummed propagator
of the gluon~\cite{LThakur, ShiChaoYi}.mainly study of the heavy quarkonium binding energy in details in~\cite{LThakur16, VKAgotiya} 
and the chemical potential effect has also been studied by the methods of dissipative hydrodynamic on quark-gluon plasma, production of photon in QGP and the quark-gluon plasma thermodynamical properties~\cite{HGervais, AMonnai, SSSingh, VSFilinov, SMSanches}.
The Alberico et al.~\cite{WMAlberico}, Mocsy et al.~\cite{AMocsy} and Agotiya et al.~\cite{VAgotiya} have solved the Schrodinger equation for the quarkonium states at finite temperature, using a temperature dependent effective potential by the linear combination of internal energy and conclude the spectral function of quarkonium in a quark-gluon plasma.
In the present work we have used the (AEIM) Analytical exact iteration method to solve the N-dimensional radial Schrodinger equation to 
investigate the properties of Quarkonium such as Bottomonium and Charmonium. 
The main aim of the present work is to find the solution of $N$-dimensional Schrodinger equation with the real part of complex valued 
potential~\cite{A.Dumitru, MatthewMargotta} at fixed value of chemical potential and finite values of temperature with anisotropic values. 
An attempt has been made to deduce the exact energy eigen value using AEIM method for calculating
the behaviour of mass spectra, binding energy and  the Dissociation temperature of quarkonium. In this paper we study the properties 
of heavy quarkonium (i.e,charmonium and bottomonium) using the anisotropic plasma in oblate case only.
The study of the relativistic heavy ion collision in the large hadron collider(LHC) at CERN and Brookhaven National Laboratory(BNL) in USA for understnding the
theory  of quarkonium bound states has been looked forward to progress significantly. 
We have studied the properties of the quarkonia states in the QGP plasma which exhibits the anisotropy in momentum space, mainly anisotropy  arises because of the
 hydrodynamical  expansion of the plasma with the non-vanishing shear viscosity. It leads to an angular 
dependence of the quarkonia potential~\cite{BKPatra}.
The dissociation of quarkonia in the quasi-particle approach for the isotropic $(\mbox{\boldmath$\xi$}=0)$ medium  has been studied in ~\cite{VKAgotiya},
the dissociation temperature$(T_D)$ of quarkonia is estimated by the local momentum anisotropy.
After considering the momentum anisotropy, we consider the oblate$(\mbox{\boldmath$\xi$}>0)$ case and compared it with the isotropic$(\mbox{\boldmath$\xi$}=0)$ case.
The main motivation to include anisotropy ($\mbox{\boldmath$\xi$}$) for the study of the properties of charmonium and botommonium
comes by the fact that the production of quark gluon-plasma in the collision of heavy-ions does not 
possess $\mbox{\boldmath$\xi$}=0$, but in the momentum-anisotropy it is present in the all dimensions 
of the direction of the stages of the off central (heavy-ion) collision.
The effect of anisotropy will significantly revise the values of $T_D$, the values of $T_D$ has observed to be lower in the oblate 
case$(\mbox{\boldmath$\xi$}>0)$ then the other cases i.e $(\mbox{\boldmath$\xi$}<0)$ and $(\mbox{\boldmath$\xi$}=0)$.
We also know that the ground state is tightly bound and hence higher values of binding energy(BE/$E_b$) and ground state  melts later than the other 
excited state. 
The paper is organised as follows. In Section II, we calculate the Exact Solution of the $N$-dimensional Radial Schrodinger equation 
with the real part of complex valued potential. In section III, we consider the debye mass from a quasiparticle picture of hot QCD and in 
section IV, we calculate the binding energy of quarkonium state in $N$-dimensional space. In section V, we calculate the mass spectra
of quarkonium state in $N$-dimensional space and finally, we conclude the work in Sec. VI.  

\begin{figure*}
    \vspace{3mm}   
    \includegraphics[height=4.6cm,width=4.6cm]{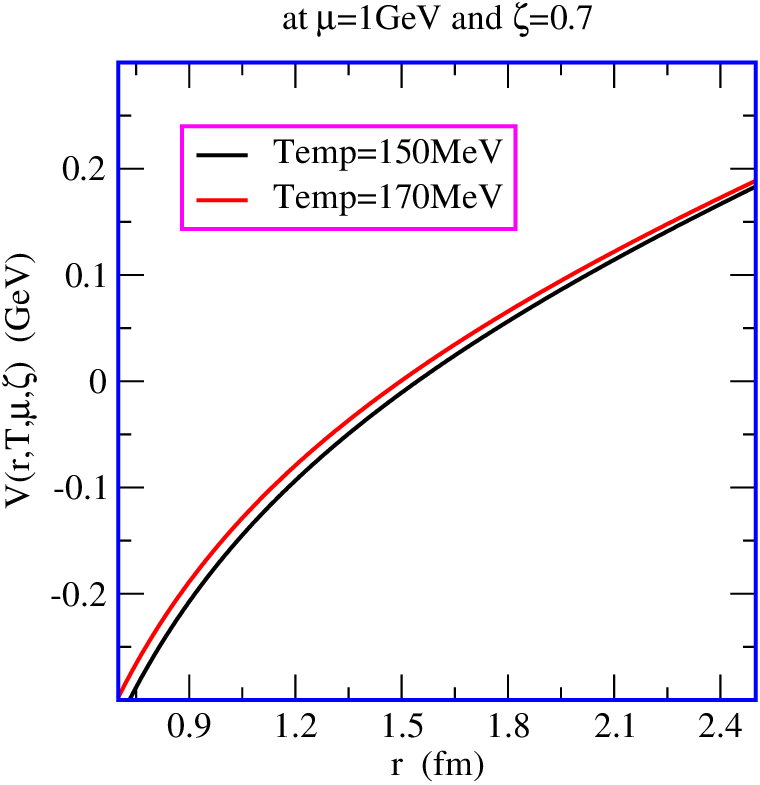}
    \hspace{2mm}
    \includegraphics[height=4.6cm,width=4.6cm]{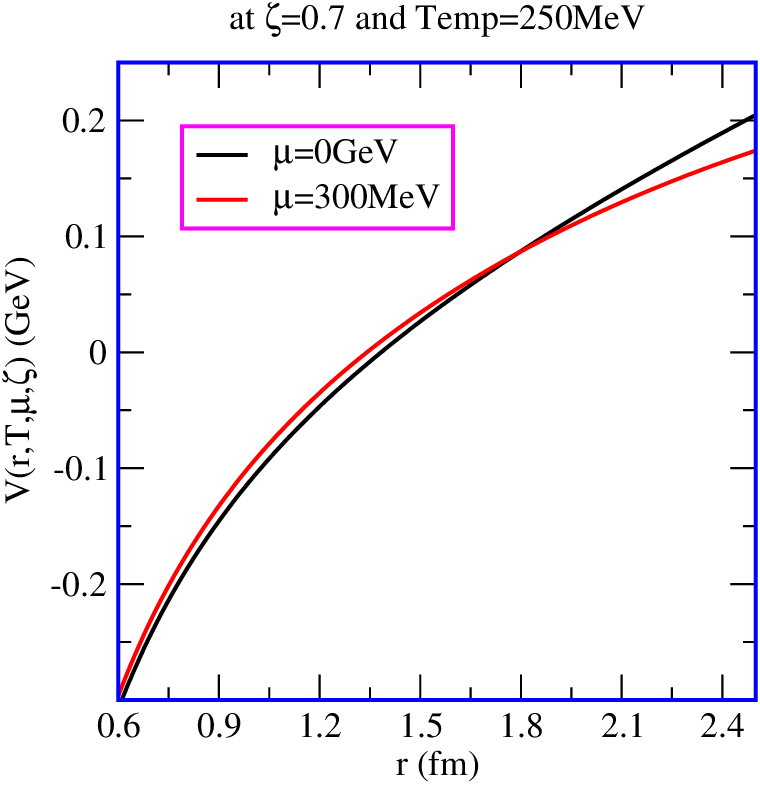} 
     \hspace{2mm}   
    \includegraphics[height=4.6cm,width=4.6cm]{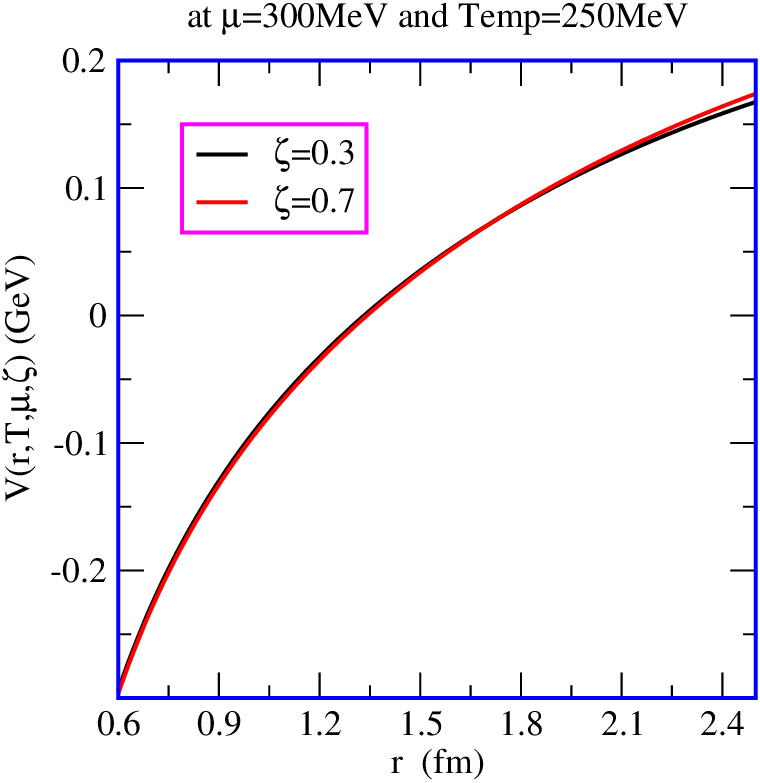}
     \vspace{3mm}  
    \caption{The variation of potential with distance(r) for the different values of the temperature (left pannel),for the different values of baryonic 
    chemical potential (middle pannel) and for the different values of anisotropy but the value of baryonic chemical potential and temperature is fixed(right pannel).}
\label{fig.1}
 \vspace{55mm}  
\end{figure*}

\section{The Solution of $N$-dimensional Radial Schrodinger equation with the real part of complex valued potential.}

The $N$-dimensional radial Schrödinger equation for two particles interacting via real part of complex valued potential takes 
the form~\cite{A.Dumitru, MatthewMargotta}

\begin{multline}
\label{eq1}
\frac{\partial^2}{\partial r^2}\psi(r)+\frac{N-1}{r}\frac{\partial}{\partial r}\psi(r)-\left(\frac{l(l+N-2)}{r^2}\right)\psi(r)\\+2\mu_{Q\bar{Q}}\left[E_{nl}-V(r)\right]\psi(r)=0
\end{multline}
where $n$,$l$ and $\mu_{Q\bar{Q}}$ are the dimensional number, angular quantum number and reduced mass of two particles respectively.
Now, the value of
\begin{equation}
\label{eq2} 
\psi (r)= R(r)/r^{\frac{N-1}{2}} 
\end{equation}
is introduced in the Eq.(\ref{eq1})equation, then we get
\begin{equation}
\label{eq3}
\left [\frac{\partial^2 }{\partial r^2}-\frac{\lambda ^2-\frac{1}{4}}{r^2}+\left (E_{nl}-V(r) \right ) \right ]R(r)=0
\end{equation}
with
\begin{equation}
\label{eq4} 
\lambda =l+\left(\frac{N-2}{2}\right)
\end{equation}
And the real part of complex valued potential $V(r)$ ~\cite{A.Dumitru} is,
\begin{multline}
\label{eq5}
V(r)=-\frac{\alpha }{r}(1+\mu r)exp(-\mu r)+\frac{2\sigma }{\mu}[1-exp(-\mu r)]\\-\sigma r exp[(-\mu r)]-\frac{0.8 \sigma }{m_{Q}^{r}}
\end{multline}
And the value of $\mu$ in the Eq.(\ref{eq5}) is,
\begin{equation}
\label{eq6}
\mu = m_{D}\left [ 1-\xi \left ( \frac{3+cos2\Theta }{16} \right ) \right ]
\end{equation}
In Eq.(\ref{eq6}) we consider the values of theeta is in parallel condition only for the simplification of the calculation.
Using exponential formula $e^{-\mu r} = \sum_{k=0 }^{\infty }\frac{(-\mu r)^{k}}{k!}$ for solving the  Eq.(\ref{eq5}) and neglect the
higher orders terms such that $\mu r \ll 1$ and then Eq.(\ref{eq5}) becomes,
\begin{equation}
\label{eq7}
V(r,\mu)=-ar^{2}+br-\frac{d}{r}
\end{equation}
where, the values of $a$,$b$ and $d$ are,
$a=\frac{\alpha \mu^{3}}{2}$,$b=\frac{\alpha \mu^{2}}{2}+\sigma$ an Eq.(\ref{eq3}) $d=\alpha +\frac{0.8 \sigma }{m_{Q}^{2}}$
Substituting in Eq.(\ref{eq8}), and after solving we get
\begin{equation}
\label{eq8}
{R}''(r)=\left [ -\varepsilon _{nl}+a_{1}r^{2}+b_{1}r^{2}-\frac{d_{1}}{r}+\frac{\lambda^{2}-\frac{1}{4} }{r^{2}} \right ]R(r)
\end{equation}
where,
$\varepsilon _{nl}=2\mu_{Q\bar{Q}}E_{nl}$, $a_{1}=\left | -2\mu_{Q\bar{Q}} a\right |$, $b_{1}=2\mu_{Q\bar{Q}}b$ and $d_{1}=2\mu_{Q\bar{Q}}d$  
The analytical exact iteration method (AEIM) requires making the following ansatz for the wave function as in~\cite{AOBarut, Sozcelik, SMIkhdair}.
\begin{equation}
\label{eq9}
R(r)=f_{n}(r)exp[g_{1}(r)]
\end{equation}
where
\begin{equation}
\label{eq10}
f_{n}(r)=1 
\end{equation}if n=0 
and
\begin{equation}
\label{eq11}
f_{n}(r)=\prod_{i=1}^{n}\left ( r-\alpha_{i}^{(n)}  \right ) 
\end{equation} for n=1,2,3........
\begin{equation}
\label{eq12}
g_{1}(r)=-\frac{1}{2}\alpha r^{2}-\beta r+\delta lnr , \alpha >0, \beta >0
\end{equation}
From Eq.(\ref{eq9}), we get
\begin{equation}
\label{eq13}
{R}''_{nl}(r)=\left ( {g}''_{l}(r)+ g_{l}^{'2}(r)+\frac{{f}''_{n}(r)+2g_{l}^{'}(r)f_{n}^{'}(r)}{f_{n}(r)} \right )R_{nl}(r)
\end{equation}
Now compare the Eq.(\ref{eq13}) and Eq.(\ref{eq9}) yields
\begin{multline}
\label{eq14}
-\varepsilon _{nl}+a_{1}r^{2}+b_{1}r^{2}-\frac{d_{1}}{r}+\frac{\lambda^{2}-\frac{1}{4} }{r^{2}}\\
= {g}''_{l}(r)+ g_{l}^{'2}(r)+\frac{{f}''_{n}(r)+2g_{l}^{'}(r)f_{n}^{'}(r)}{f_{n}(r)}
\end{multline}
at $n=0$, substitute Eqs.(\ref{eq10}),(\ref{eq11}),(\ref{eq11}) and Eq.(\ref{eq12}) into Eq.(\ref{eq14})we get
\begin{multline}
\label{eq15}
a_{1}r^{2}+b_{1}r-\frac{d_{1}}{r}+\frac{\lambda^{2}-\frac{1}{4} }{r^{2}}-\varepsilon_{0l}
= \alpha^{2}r^{2}+2\alpha \beta r-\alpha [1+2(\delta +0)]\\+\beta ^{2}-\frac{2\beta \delta }{r}+\frac{\delta (\delta -1)}{r^{2}}
\end{multline}
Now comparing the corresponding power of $r$ on both side of Eq.(\ref{eq15}) then obtains
\begin{equation}
\label{eq16}
\alpha =\sqrt{a_{1}}
\end{equation}
\begin{equation}
\label{eq17}
\beta = \frac{b_{1}}{2\sqrt{a_{1}}}
\end{equation}
\begin{equation}
\label{eq18}
d_{1}=2\beta (\delta +0)
\end{equation}
\begin{equation}
\label{eq19}
\delta = \frac{1}{2}(1\pm 2\lambda )
\end{equation}
\begin{equation}
\label{eq20}
\varepsilon_{0l} =\alpha [1+2(\delta +0)]-\beta ^{2}
\end{equation}
Now, the energy eigen value formula for the ground state is:
\begin{equation}
\label{eq21}
E_{0l}= \sqrt{\frac{a}{2\mu_{Q\bar{Q}}}}(N+2l)-\frac{b^{2}}{4a}
\end{equation}
Now for the first node (n=$1$), we used the function 
$f_{1}(r)=\left ( r-\alpha_{1}^{(1)}  \right )$ and $g_{1}(r)$ then
\begin{multline}
\label{eq22}
a_{1}r^{2}+b_{1}r-\frac{d_{1}}{r}+\frac{\lambda^{2}-\frac{1}{4} }{r^{2}}-\varepsilon_{1l} = 
\alpha ^{2}r^{2}+2\alpha \beta r-\alpha [1+2(\delta +1)]\\+\beta^{2}-\frac{2[\beta (\delta +1)
+\alpha \alpha_{1}^{(1)} ]}{r}+\frac{\delta (\delta -1)}{r^{2}}
\end{multline}
after comparing the coefficients of $r$, the relation between the potential parameters are:
\begin{equation}
\label{eq23}
\alpha =\sqrt{a_{1}}
\end{equation}
\begin{equation}
\label{eq24}
\beta =\frac{b_{1}}{2\sqrt{a_{1}}}
\end{equation}
\begin{equation}
\label{eq25}
d_{1}=2\beta (\delta +1)
\end{equation}
\begin{equation}
\label{eq26}
\delta =\frac{1}{2}(1\pm 2\lambda )
\end{equation}
\begin{equation}
\label{eq27}
\varepsilon_{1l} = \alpha [1+2(\delta +1)]-\beta^{2}
\end{equation}
\begin{equation}
\label{eq28} 
d_{1}-2\beta (\delta +1)=2\alpha \alpha_{1}^{(1)}
\end{equation}
\begin{equation}
\label{eq29}
(d_{1}-2\beta \delta )\alpha_{1}^{(1)} =2\delta 
\end{equation}
now the energy eigen value formula for first excited state $E_{1l}$ is:
\begin{equation}
\label{eq30}
E_{1l}=\sqrt{\frac{a}{2\mu_{Q\bar{Q}}}}(N+2l+2)-\frac{b^{2}}{4a}
\end{equation}
Similarly for second node $(n=2)$, we use $f_{2}(r)=\left ( r-\alpha _{1}^{(2)} \right ) \left ( r-\alpha _{2}^{(2)} \right )$ and $g_{1}(r)$
and we get,
\begin{multline}
\label{eq31}
a_{1}r^{2}+b_{1}r-\frac{d_{1}}{r}+\frac{\lambda^{2}-\frac{1}{4} }{r^{2}}-\varepsilon _{2l}=\alpha^{2} r^{2}+2\alpha \beta r
-\alpha [1+2(\delta +2)]\\+\beta ^{2}-\frac{2[\beta (\delta +2)+\alpha (\alpha_{1}^{(2)}+\alpha_{2}^{(2)}  )]}{r}+\frac{\delta (\delta -1)}{r^{2}}
\end{multline}
after comparing the coeficients the relation is:
\begin{equation}
\label{eq32}
\alpha =\sqrt{a_{1}}
\end{equation}
\begin{equation}
\label{eq33}
\beta =\frac{b_{1}}{2\sqrt{a_{1}}}
\end{equation}
\begin{equation}
\label{eq34}
\delta =\frac{1}{2}(1\pm 2\lambda )
\end{equation}
\begin{equation}
\label{eq35}
\varepsilon_{2l}=\alpha [1+2(\delta +2)]-\beta^{2}
\end{equation}
\begin{equation}
\label{eq36} 
d_{1}-2\beta (\delta +2)=2\alpha (\alpha_{1}^{(2)}+\alpha_{2}^{(2)}  )
\end{equation}
\begin{equation}
\label{eq37}
(d_{1}-2\beta \delta)\alpha_{1}^{(2)} \alpha_{2}^{(2)}=2\delta (\alpha_{1}^{(2)}+\alpha_{2}^{(2)})
\end{equation}
\begin{equation}
\label{eq38}
{d_{1}-2\beta (\delta +1)}(\alpha_{1}^{(2)}+\alpha_{2}^{(2)}  )=4\alpha(\alpha_{1}^{(2)} \alpha_{2}^{(2)})+2(2\delta +1)
\end{equation}
The energy eigen value formula for $E_{2l}$ is
\begin{equation}
\label{eq39}
E_{2l}=\sqrt{\frac{a}{2\mu_{Q\bar{Q}}}}(N+2l+4)-\frac{b^{2}}{4a}
\end{equation}
Hence, Iteration method is repeated similarly many times. therefore, exact energy eigen value for the state of 
finite temperature and chemical baryonic potential also with anisotropy in $N$-dimensional space is:
\begin{equation}
\label{eq40}
E_{nl}^{n}=\sqrt{\frac{a}{2\mu_{Q\bar{Q}}}}(N+2l+2n)-\frac{b^{2}}{4a}
\end{equation} , $n=0,1,2,3........$.
\begin{figure*}
    \vspace{2mm}   
    \includegraphics[height=5.6cm,width=5.6cm]{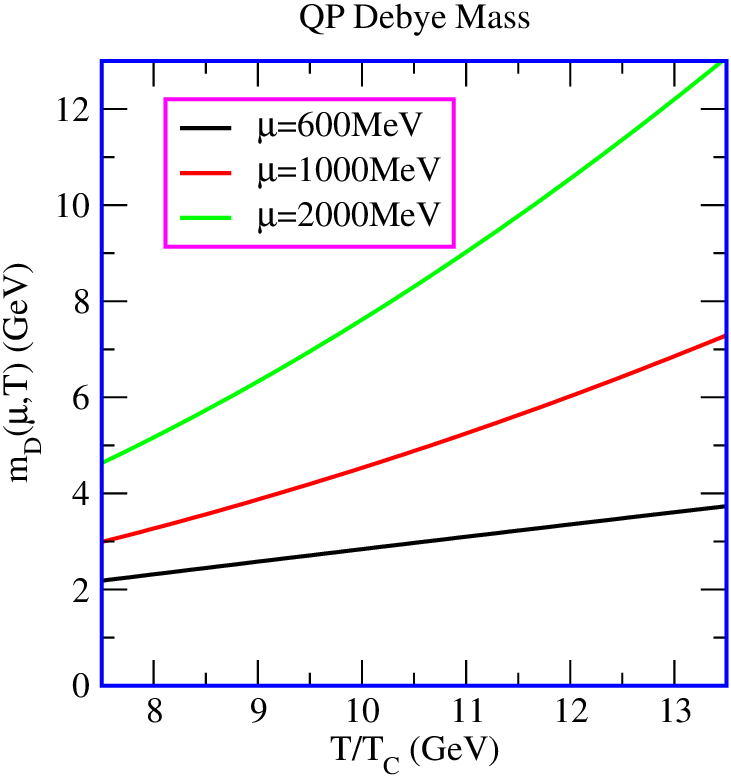}
    \hspace{6mm}
    \includegraphics[height=5.6cm,width=5.6cm]{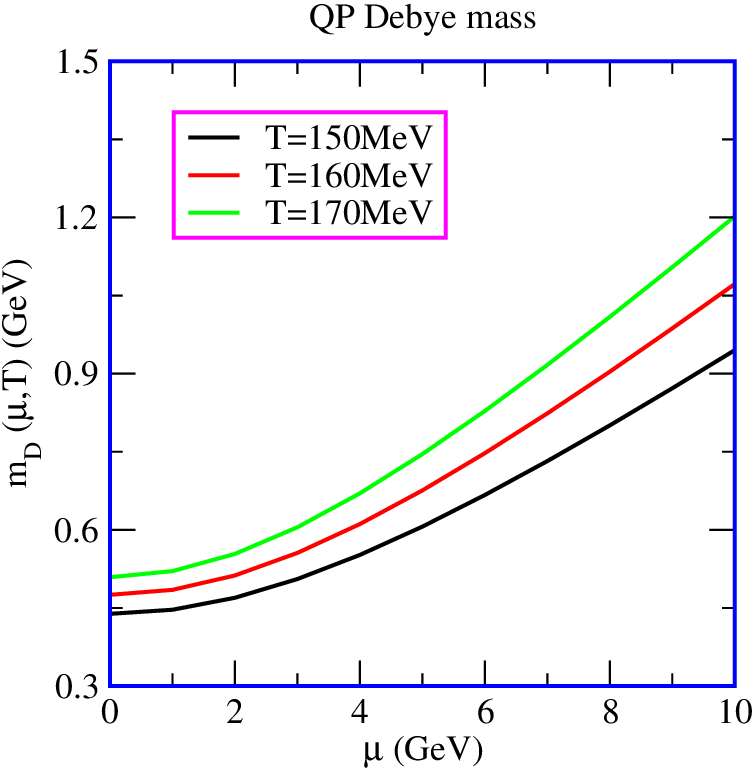}
\caption{The variation of quasi-particle debye mass with temperature at different values of baryonic chemical potential(left pannel) 
and with baryonic chemical potential at different values of temperature(right pannel).}
\vspace{8mm}
\label{fig.2}
\end{figure*}

\section{ The Debye mass from a quasiparticle picture of hot QCD}

The Debye mass in leading order in QCD coupling at high temperature has been known for a long time and is perturbative in nature.
Rebhan~\cite{ARebhan} has defined the Debye mass $m_D$ by seeing the pole of the static propagator which is relevant instead of 
the time-time component of the gluon self energy and obtained a Debye mass which is gauge independent. This result follows from
the fact that the pole of self energy does not depend on the choice of the gauge. The Debye mass was calculated for QGP at a high 
temperature in next to lead order (NLO) in QCD coupling from correlation of two polyakov loop by Braaten and Nieta~\cite{EBraaten}, 
this result agrees with the HTL result~\cite{ARebhan}. It was pointed out by Arnold and Yaffe~\cite{PArnold} that the physics of
confined magnetic charge was to be known in order to understand the contribution of O$(g^2 T)$ to the Debye mass in QCD, it is also
pointed out by them that the Debye mass as a pole of gluon propagator no longer holds true. Importantly in lattice QCD, the definition 
of Debye mass its self encounter difficulty due to the fact that unlike QED the electric field correlators are not gauge invariant in QCD. 
The proposal of this problem is based on effective theories obtained by dimensional reduction~\cite{KKajantie}, spatial correlation 
function of gauge-invariant meson energy and the behavior of color singlet free energies~\cite{SNadkarni} has been made Burnier and 
Rothkopf~\cite{YBurnier} has attempted to defined a gauge invariant mass from a complex static in medium heavy-quark potential obtained from lattice QCD.
\begin{figure*}
    \vspace{25mm}   
    \includegraphics[height=6.6cm,width=6.6cm]{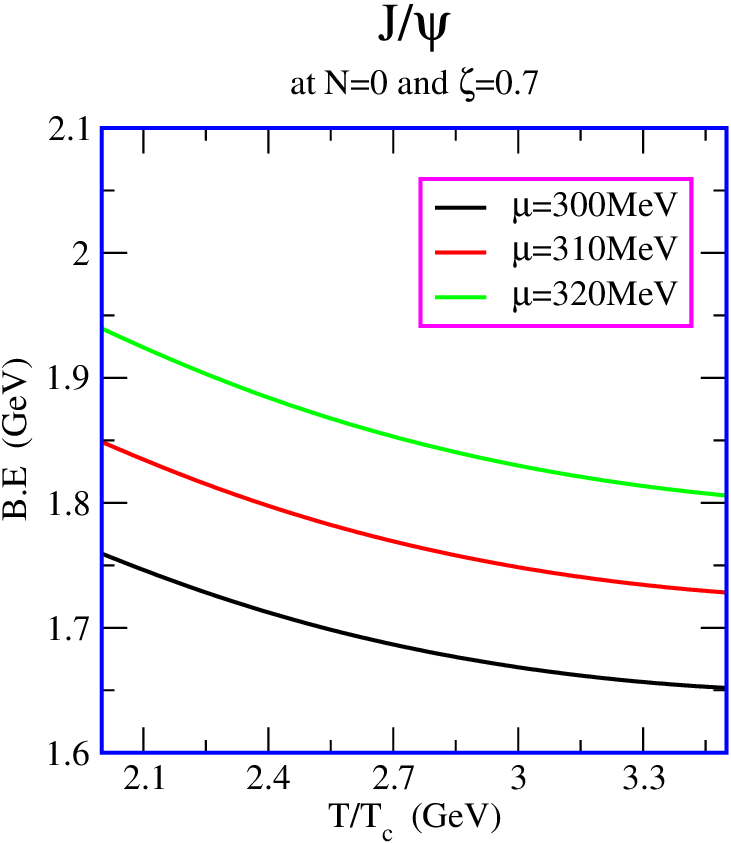}
    \hspace{7mm}
    \includegraphics[height=6.6cm,width=6.6cm]{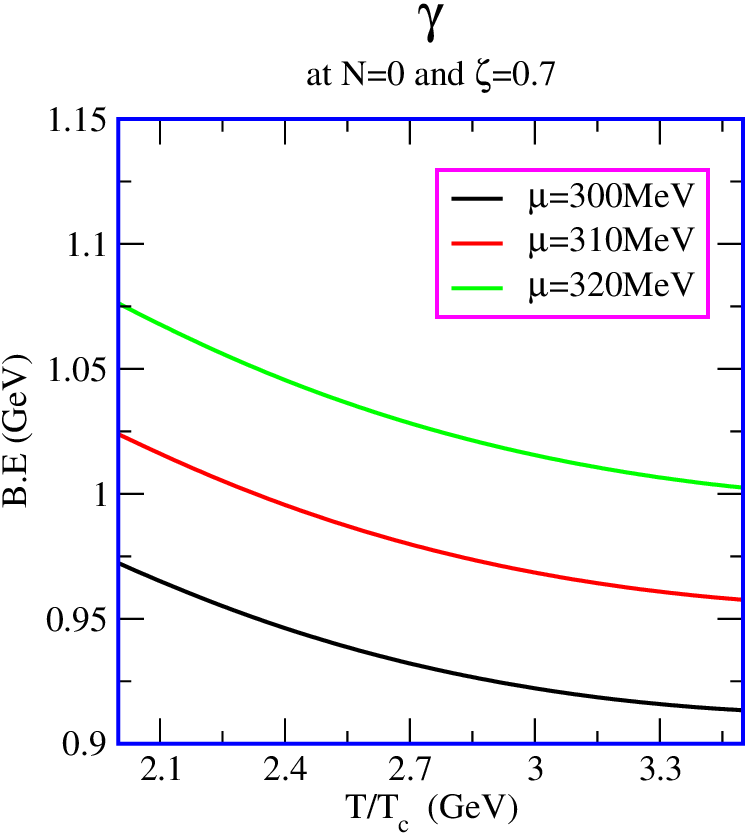}
    \vspace{5mm} 
\caption{Dependence of $J/\psi$ binding energy with temperature (left pannel) and dependence of $\Upsilon$ binding energy with temperature
(right pannel) at different values of baryonic chemical potential but the value of $N$ and \mbox{\boldmath$\xi$} is fixed.}
\label{fig.3}
 \vspace{25mm} 
\end{figure*}
\begin{figure*}
    \vspace{3mm}   
    \includegraphics[height=6.4cm,width=6.4cm]{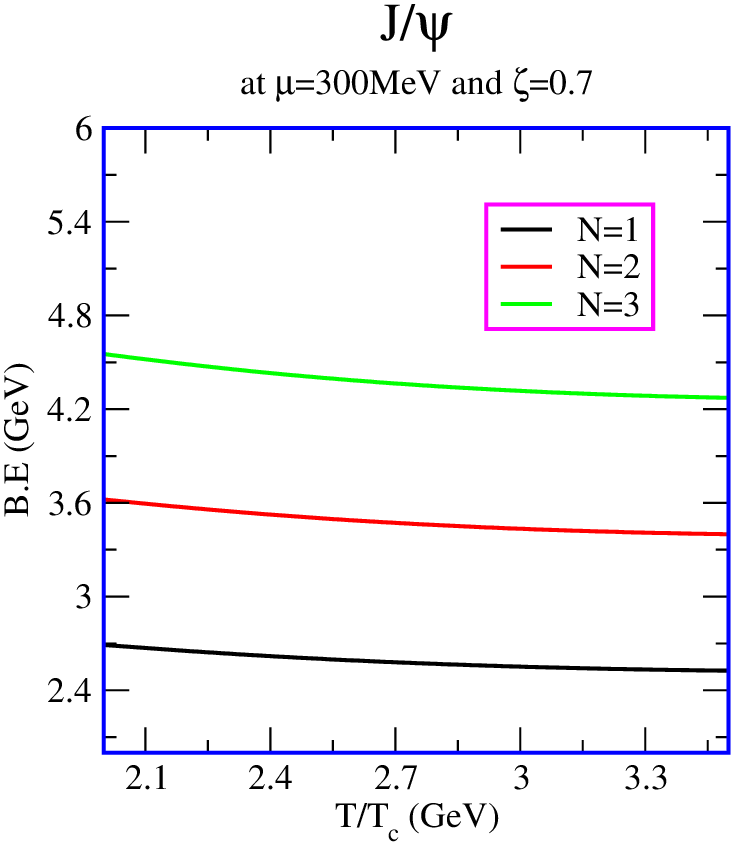}
    \hspace{7mm}
    \includegraphics[height=6.4cm,width=6.4cm]{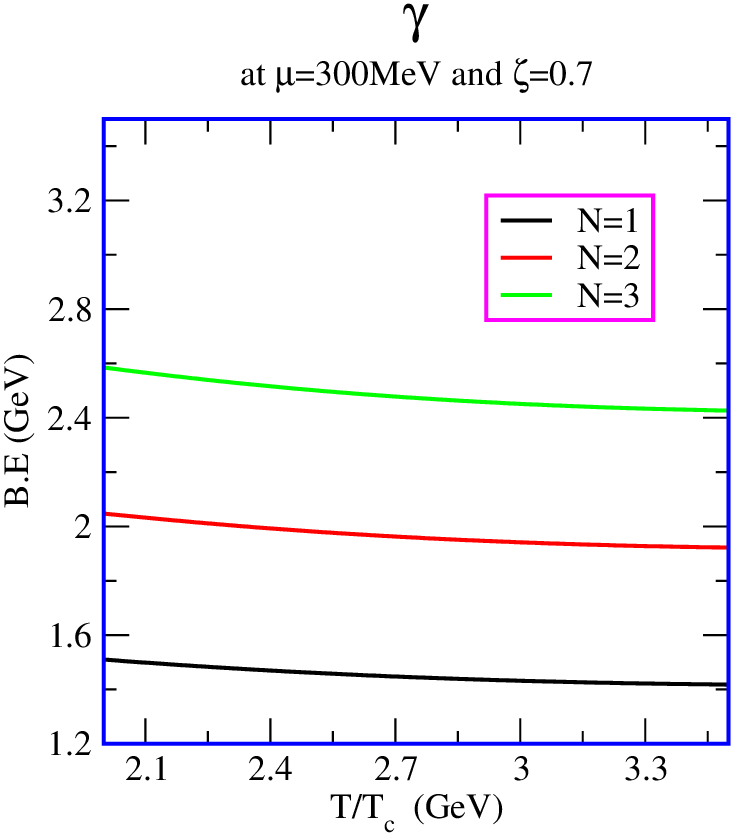}
\caption{Dependence of $J/\psi$ binding energy with temperature (left pannel) and dependence of $\Upsilon$ binding energy with
temperature (right pannel) at different values of dimensionality number but the value of baryonic chemical potential and \mbox{\boldmath$\xi$}  is fixed.}
\label{fig.4}
 \vspace{55mm} 
\end{figure*}
Several attempts has been made to capture all the interaction effects present in hot QCD equation of state (EOS) in terms of non-interacting 
quasi-partons (quasi-gluons and quasi-quarks). These quasiparton are the excitations of the interacting quarks and gluons and they can be
categorized as, effective mass model~\cite{VGoloviznin, peshier}, effective mass model with polykov loop~\cite{MDElia}, model based on 
PNJL and NJL~\cite{ADumitru} and $(4)$ effective fugacity model~\cite{VChandra7, VChandra9}. In QCD the quasipartical model is a
phenomenological model which is widely used to describe the nonlinear behaviour of QGP near phase transition point. In this model a system of
interacting massless quarks and gluon can be described as an ideal gas of "massive" non interacting quasiparticle. The mass of the quasiparticle
is dependent on the temperature which arises due to the interaction of gluons and quarks with surrounding medium. The quasiparticle retain the quantum
number of the quarks and gluons~\cite{PKSrivastava}.
\begin{figure*}
    \vspace{3mm}   
    \includegraphics[height=6.4cm,width=6.4cm]{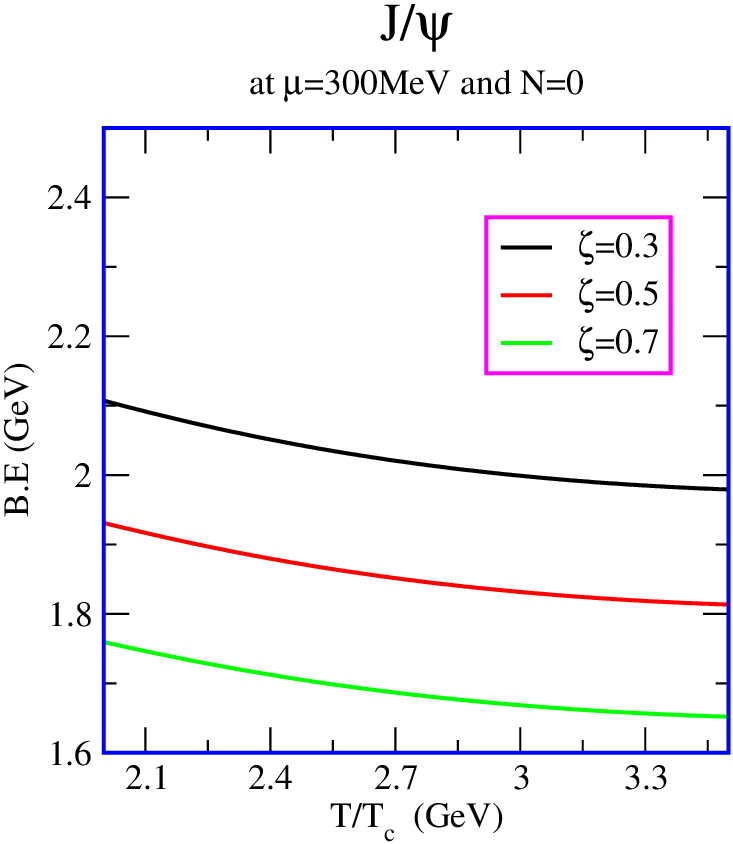}
    \hspace{7mm}
    \includegraphics[height=6.4cm,width=6.4cm]{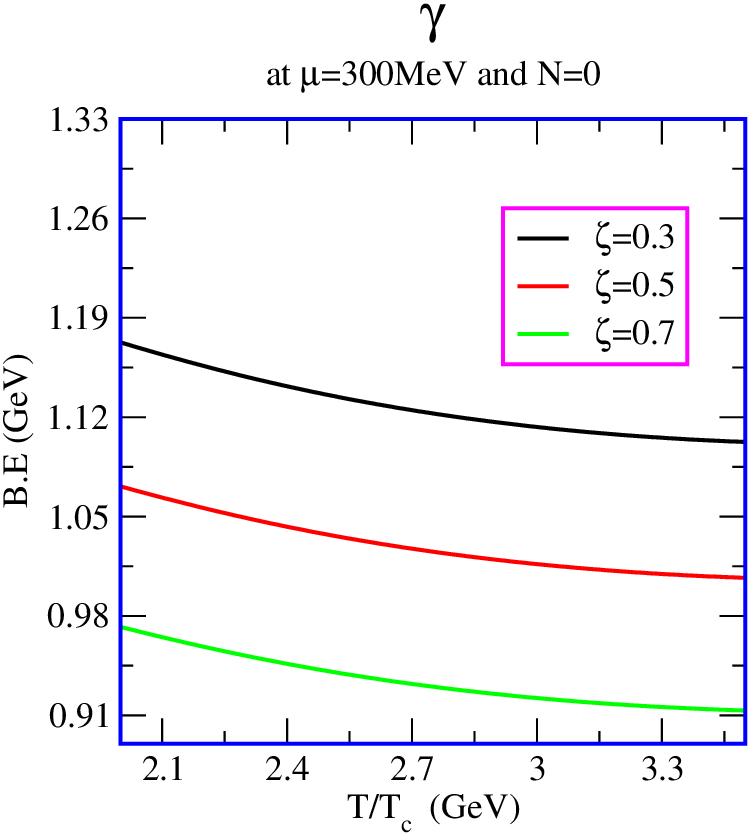}
\caption{Dependence of $J/\psi$ binding energy with temperature (left pannel) and dependence of $\Upsilon$ binding energy with
temperature (right pannel) at different values of \mbox{\boldmath$\xi$} but the value of $\mu$ and $N$ is fixed.}
\label{fig.5}
 \vspace{48mm} 
\end{figure*}
\begin{figure*}
    \vspace{3mm}   
    \includegraphics[height=6.4cm,width=6.4cm]{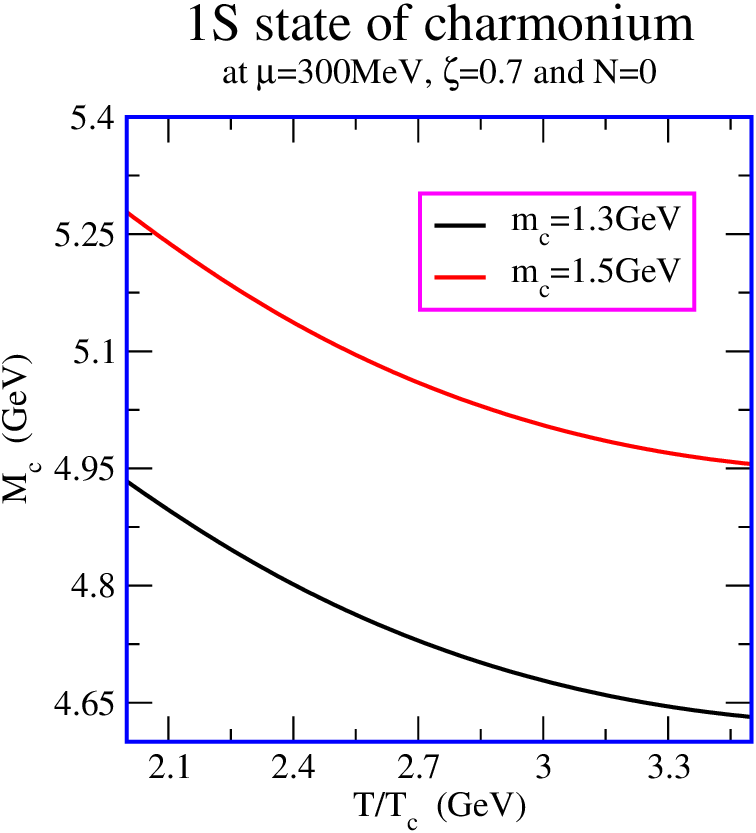}
    \hspace{7mm}
    \includegraphics[height=6.4cm,width=6.4cm]{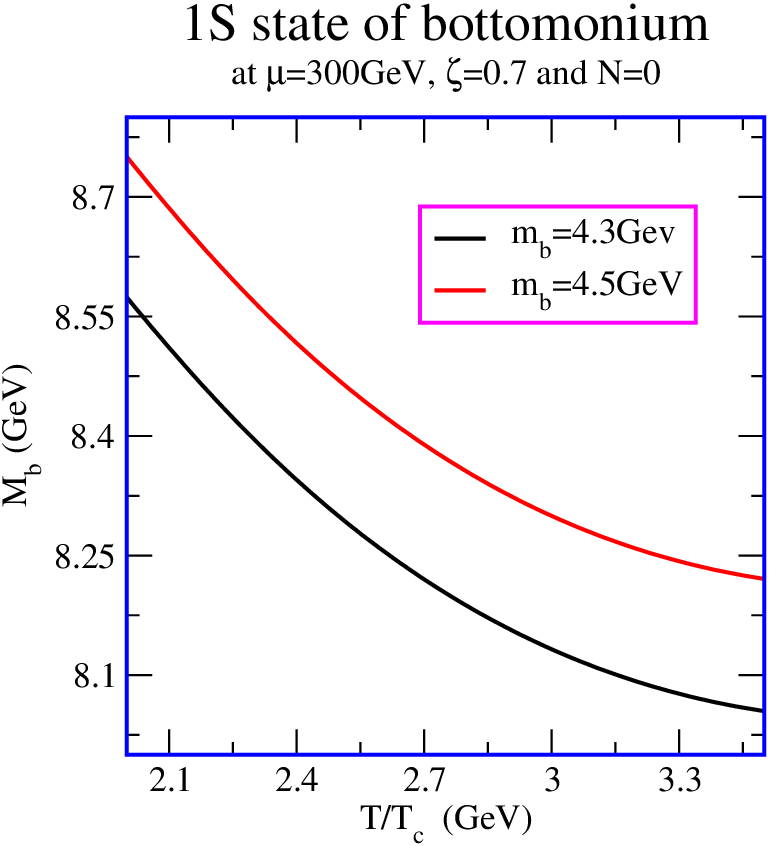}
\caption{Dependence of mass of charmonium with temperature (left pannel) and dependence of mass of bottomonium with temperature (right pannel).}
\label{fig.6}
 \vspace{25mm} 
\end{figure*}
Here, ${m_D}$ is the quasi-particle Debye mass. In our calculation, we use the Debye mass $m_D$ in the pure gluonic case as: 
\begin{equation}
\label{eq41}
m^2_D=g^2(T) T^2 \bigg(\frac{N_c}{3}\times\frac{6 PolyLog[2,z_g]}{\pi^2}\bigg)
\end{equation}
and ${m_D}$ for full QCD case is:
\begin{eqnarray}
\label{eq42}
m^2_D\left(T,\mu_f\right) &=& g^2(T) T^2 \bigg[
\bigg(\frac{N_c}{3}\times\frac{6 PolyLog[2,z_g]}{\pi^2}\bigg)\nonumber\\&&
+{\bigg(\frac{\hat{N_f}}{6}\times\frac{-12 PolyLog[2,-z_q]}{\pi^2}\bigg]}
\end{eqnarray}
and 
\begin{eqnarray}
\label{eq43}
\hat{N_f} &=& \bigg(N_f +\frac{3}{\pi^2}\sum\frac{\mu^2}{T^2}\bigg)
\end{eqnarray}
Here, $g(T)$ is the QCD running coupling constant, $N_c=3$ ($SU(3)$) and $N_f$ is the number of flavor,
the function $PolyLog[2,z]$ having form, $PolyLog[2,z]=\sum_{k=1}^{\infty} \frac{z^k}{k^2}$ and $z_g$ is the
quasi-gluon effective fugacity and $z_q$ is quasi-quark effective fugacity. These distribution functions are
isotropic in nature,
\begin{eqnarray}
\label{eq44}
f_{g,q}=\frac{z_{g,q}exp(-\beta p)}{\left (1\pm z_{g,q}exp(-\beta p)  \right )}
\end{eqnarray}
\begin{figure*}
    \vspace{3mm}   
    \includegraphics[height=6.4cm,width=6.4cm]{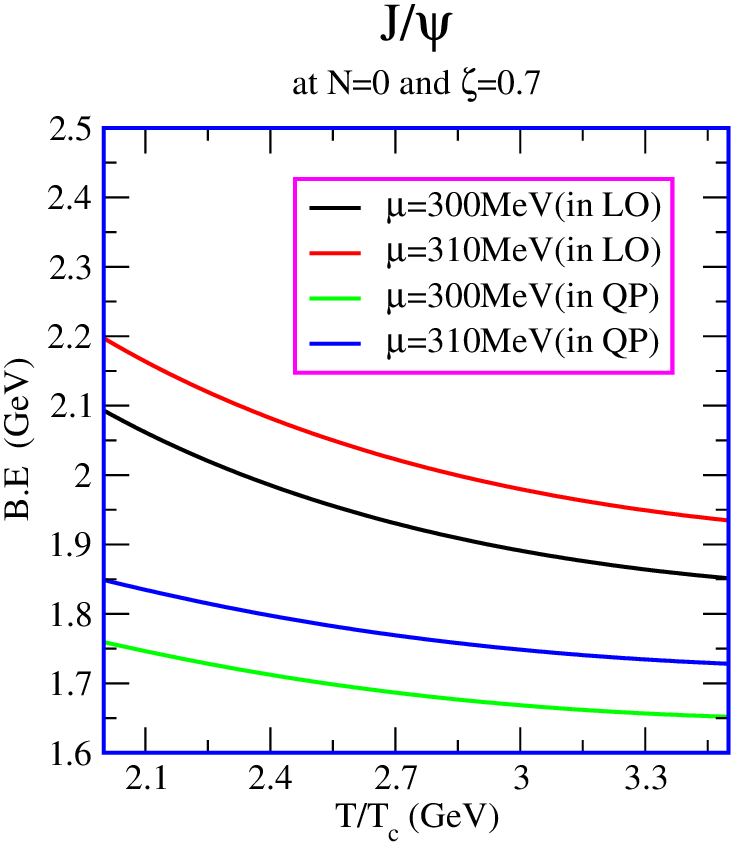}
    \hspace{7mm}
    \includegraphics[height=6.4cm,width=6.4cm]{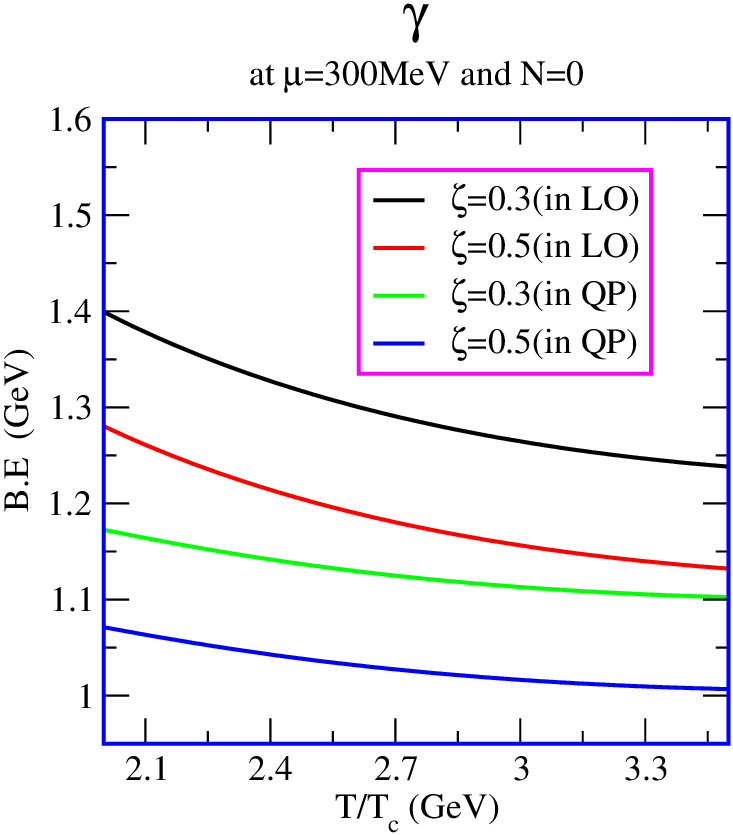}
\caption{Dependence of $J/\psi$ binding energy with temperature (left pannel) at different values of $\mu$ and dependence of $\Upsilon$ binding energy with
temperature (right pannel) at different values of \mbox{\boldmath$\xi$}.}
\label{fig.7}
 \vspace{25mm} 
\end{figure*}
where, $g$ define the study for quasi-gluons and $q$ defines for quasi-quarks.
These fugacities should not be confused with any conservations law (number conservation) 
and have merely been introduced to encode all the interaction effects at high temperature QCD. Both $z_g$ and $z_q$ 
have a very complicated temperature dependence and asymptotically reach to the ideal value unity~\cite{VChandra9}.
The temperature dependence $z_g$ and $z_q$  fits well to the form given below,
\begin {equation}
\label{eq45}
z_{g,q}=a_{q,g}\exp\bigg(-\frac{b_{g,q}}{x^2}-\frac{c_{g,q}}{x^4}-\frac{d_{g,q}}{x^6}\bigg).
\end {equation}
(Here $x=T/T_c$ and $a$, $b$ and $c$ and $d$ are fitting parameters), for both EOS$1$ and EOS$2$. Here, EoS$1$ is 
the $O(g^5)$ hot QCD~\cite{PArnold} and EoS$2$ is the $O(g^6\ln(1/g)$ hot QCD EoS~\cite{KKajantie} in the 
quasi-particle description~\cite{VChandra7, VChandra9} respectively.
Now, the expressions for the Debye mass can be rewritten in terms effective charges for the quasi-gluons and quarks for pure gauge as:
\begin{equation}
\label{eq46}
m^2_D\left(T,\mu\right)= Q^2_g T^2\frac{N_c}{3}
\end{equation}
and after introducing Eq.(\ref{eq44}) in Eq.(\ref{eq43}), then the expression of full QCD equation becomes:
\begin{multline}
\label{eq47}
m^2_D\left(T,\mu\right)=T^2( \frac{N_c}{3} Q^2_g)+\bigg(\bigg(\frac{N_f}{6}+\frac{1}{2\pi^2}\bigg(\frac{\mu^2}{T^2}\bigg)\bigg)Q^2_q\bigg)
\end{multline}
And in the Eq.(\ref{eq48}) the $\mu$ is defined as:
\begin{eqnarray}
\label{eq48}
\mu &=& \frac{\mu_b}{3}
\end{eqnarray}
where, $(\mu)$ defined the quark-chemical potential and $(\mu_b)$ is baryonic chemical potential.
where, $Q_g$ and $Q_q$ are the effective charges given by the equations:
\begin{eqnarray}
\label{eq49}
Q^2_g&=&g^2 (T) \frac{6 PolyLog[2,z_g]}{\pi^2}\nonumber\\
Q^2_q&=&g^2 (T)  \frac{-12 PolyLog[2,-z_q]}{\pi^2}.
\end{eqnarray}
 In our present analysis we had used the temperature dependence of the quasi-particle Debye mass, $m_D^{QP}$ in full QCD with $N_f=3$ and to
 determine the temperature dependence of binding energy and dissociation temperature.

\section{Binding energy of Quarkonium state in $N-$ dimensional space}

The binding energy of quarkonium state such as Bottomonium and Charmonium is studied in this section. We have calculated by 
the $N-$dimensional space at fixed value of $\mu$, at finite temperature and also with the values of isotropy $({\mbox{\boldmath$\xi$}=0})$ and anisotropy.
After using iteration method, we found the exact energy formula in $N-$dimensional space is:
\begin{equation}
\label{eq50}
E_{nl}^{n}=\sqrt{\frac{a}{2\mu_{Q\bar{Q}}}}(N+2l+2n)-\frac{b^{2}}{4a}
\end{equation} 
Substituting the value of n (i.e, $n=0,1,2,3....$) according to the state of quarkonia like $1$S and $2$S etc.
Now, we substitute the value of $a$ and $b$ in the above Eq.(\ref{eq51}) and we get the binding energy for the quarkonium state is
\begin{equation}
\label{eq51}
E_{b}=\sqrt{\frac{\alpha \mu^{3}}{4\mu_{Q\bar{Q}}}}(N+2l+2n)-\frac{(\frac{\alpha \mu^{2}}{2}+\sigma )^{2}}{4\times\frac{\alpha \mu^{3}}{2}}
\end{equation}
or,
\begin{equation}
\label{eq52}
E_{b}=\sqrt{\frac{\alpha \mu^{3}}{4\mu_{Q\bar{Q}}}}(N+2l+2n)-\left [ \frac{\alpha \mu}{8} + \frac{\sigma^{2} }{2\alpha \mu^{3}} + \frac{\sigma }{2\mu}\right ]
\end{equation}
where $\mu_{Q\bar{Q}}$=$\frac{m_{c}}{2}$ and $\frac{m_{b}}{2}$.
 
\section{Mass Spectra of Quarkonium state in $N$-dimensional space}

For calculating the mass spectra of heavy quarkonia the relation is:
\begin{equation}
\label{eq53}
M=2m_{Q}+E_{nl}^{n}
\end{equation}
Here, mass spectra is equal to the sum of the exact formula of energy and twice the quark-mass. Now we substitute 
the values of $E_{nl}^{n}$ from Eq.(\ref{eq51}) in the Eq.(\ref{eq54}) as,
\begin{equation}
\label{eq54}
M=2m_{Q}+\sqrt{\frac{a}{2\mu_{Q\bar{Q}}}}(N+2l+2n)-\frac{b^{2}}{4a}
\end{equation}
Now, substitute the values of $a$ and $b$ in the Eq.(\ref{eq55}) we get: 
\begin{equation}
\label{eq55}
M=2m_{Q}+\sqrt{\frac{\alpha \mu^{3}}{4\mu_{Q\bar{Q}}}}(N+2l+2n)-\frac{(\frac{\alpha \mu^{2}}{2}+\sigma )^{2}}{4\times\frac{\alpha \mu^{3}}{2}}
\end{equation}
and
\begin{equation}
\label{eq56}
M=2m_{Q}+\sqrt{\frac{\alpha \mu^{3}}{4\mu_{Q\bar{Q}}}}(N+2l+2n)-\left [ \frac{\alpha \mu}{8} + \frac{\sigma^{2} }{2\alpha \mu^{3}} + \frac{\sigma }{2\mu}\right ]
\end{equation}
where $m_{Q}$ is used for the mass of quarkonium state such as charmonium and bottomonium mass.

\section{Results and Conclusion}
\label{RD}

In the present analysis, the value of critical temperature ($T_c=197MeV$) have been fixed, and find the various quantities after considering the anisotropy values in oblate case only 
in the hot quantum-chromodynamics (QCD) plasma. 
The value of (\mbox{\boldmath$\xi$}) have been increased in the right pannel of figure\ref{fig.1}, noticed that the separation of potential near the axis of origin is
slightly increased, and same behaviour is also shown if the value of chemical potential have been increased in the figure \ref{fig.1} (middle pannel). But in figure \ref{fig.1} 
(left pannel) the separation of potential near the axis of origin is slightly decreased after increased the value of temperature. from this we conclude that there is little effect of anisotropy on the real part of the potential.
Figure \ref{fig.2}, shows the variation of quasiparticle debye mass with temperature at different values of baryonic chemical potential(left pannel) 
and with baryonic chemical potential at different values of temperature (right pannel). Where the baryonic chemical potential$(\mu_b)$ is related to the three times of the 
quark-chemical potential$(\mu)$. The screening mass at baryon density and temperature has studied by the lattice taylor's expansion method~\cite{MDoring}.
In the figure \ref{fig.2} it has been observed that with increase in the values of chemical potential, the debye mass increases (left pannel), same behaviour for the debye mass has been seen for the temperature (right pannel). from the energy eigen values obtained through N-dimensional Schrodinger equation using AEIM method, we calculate the binding energy, mass spectra and dissociation temperature. 
In figures \ref{fig.3}, \ref{fig.4} and \ref{fig.5} shows, the variation of binding energy of $J/\psi$ and $\Upsilon$ with temperature.
After increased the value of anisotropy (\mbox{\boldmath$\xi$}) in figure \ref{fig.5}, the variation of binding energy is decreased. Because 
there are mainy two cases: first one is that, the value of anisotropy have been increased, the binding energy is stronger in the comparison of isotropic and prolate case. 
Since the potential is more deep with the increases of (\mbox{\boldmath$\xi$}) because of weaker screening. The screening of the string part and coloumb 
part are less attenuated in that case the quarkonium binding energy becomes larger then the case of isotropic medium, and also there is a strong decreasing 
trend with the temperature.
It has also been noticed that, the variation of binding energy is depend upon the mass of the state, if the value of the quark-mass of the
state is increased, the binding energy decreased.
The effect of dimensionality number$(N)$ on the binding energy for
the quarkonium state $J/\psi$ and $\Upsilon$ with temperature have been shown in figure \ref{fig.4}. The value of dimensionality number have been increased then
the binding energy of the state of quarkonia is also increased at the fixed value of baryonic chemical potential($\mu$) and (\mbox{\boldmath$\xi$}).
Similarly in figure \ref{fig.3}, we also noticed that the binding energy is increased with increasing the values of chemical potential at constant value of anisotropy and dimensionality number.
But in figure \ref{fig.6}, the variation of mass spectra of heavy quarkonia with temperature for $1S$ state of charmonium (left pannel) 
and $1S$ of bottomonium (right pannel) is observed using two different value of masses of charmonium and bottomonium. If we increased the value of quark mass, the mass
spectra of heavy quarkonia is also
increased~\cite{WMAlberico}.
And the same variation of mass spectra is also observed for the excited states of quarkonia.
In figure \ref{fig.7}, we have compared the binding energy of Leading order(LO) and Quasi-Particle(QP) Debye mass with chemical potential (in left pannel) and with anisotropy (in right pannel), and it has been noticed that the value of binding energy in terms of LO debye mass is greater than that of QP debye mass. The LO Debye mass at high temperature range is known from long time and is perturbative~\cite{E.Shuryak} in nature. But in the present analysis, we consider the LO and QP debye mass in terms of chemical potential.
However, we also know that about the lattice parameterized form($m_D^L$) is $1.4$ times the LO debye mass, in that case, we observed that ($m_D^L$) the variation of binding energy of lattice parameterized form of Debye mass is greater than that of LO debye mass.
Thus the relation between binding energy and temperature is poised to give the information about the pattern of dissociation temperature($T_D$) in thermal medium, which will now be used to calculate the $T_D$ for the different quarkonium states.
Recent studies shows various methods for obtaining the dissociation temperature for different states of quarkonia.~\cite{PSandin}, authors have investigate the $T_D$ of heavy quarkonia by the thermal width.~\cite{AMocsy}, 
authors have apply a condition for calculating the $T_D$ i.e, thermal width is greater than equal to the twice of the real
part of binding energy.~\cite{VAgotiya}, authors have calculated the lower and upper bound state of the $T_D$ by applying
the condition for the dissociation i.e, $E_{bin}=3T_D$ and $E_{bin}=T_D$ respectively.~\cite{UK}, authors obtained the $T_D$ 
of the states of quarkonia when the binding energy is of the order of their chemical baryonic potential.
\begin{equation}
\label{eq57}
\sqrt{\frac{\alpha \mu^{3}}{4\mu_{Q\bar{Q}}}}(N+2l+2n)-\left [ \frac{\alpha \mu}{8} + \frac{\sigma^{2} }{2\alpha \mu^{3}} + \frac{\sigma }{2\mu}\right ] = 3(T_D)
\end{equation}
where $\mu_{Q\bar{Q}}$=$\frac{m_{c}}{2}$ and $\frac{m_{b}}{2}$.
\begin{table}
\label{table1}
\centering
\caption{The dissociation temperature(is in unit of $T_c$) of QP Debye mass for lower bound state $(T_D)$ with $T_c=197MeV$ for the different  ground states of quarkonia i.e, ${J/\psi}$ and ${\Upsilon}$ at $\mu=300MeV$ and for $N=0$.}
\vspace{4mm}
\begin{tabular}{|l|l|l|l|}
\hline
$State$ & $\mbox{\boldmath$\xi$}=0$ & $\mbox{\boldmath$\xi$}=0.3$ & $\mbox{\boldmath$\xi$}=0.7$\\
\hline         
$J/\psi$ & 3.54060 & 3.14720 & 2.65228\\
\hline
$\Upsilon$ & 2.05583 & 1.82741 & 1.53553\\
\hline
\end{tabular}
\end{table}
\begin{table}
\label{table2}
\centering
\caption{The dissociation temperature(is in unit of $T_c$) of LO Debye mass for lower bound state $(T_D)$ with $T_c=197MeV$ for the different  ground states of quarkonia i.e, ${J/\psi}$ and ${\Upsilon}$ at $\mu=300MeV$ and for $N=0$.}
\vspace{4mm}
\begin{tabular}{|l|l|l|l|}
\hline
$State$ & $\mbox{\boldmath$\xi$}=0$ & $\mbox{\boldmath$\xi$}=0.3$ & $\mbox{\boldmath$\xi$}=0.7$\\
\hline         
$J/\psi$ & 4.03553 & 3.57868 & 3.03299\\
\hline
$\Upsilon$ & 2.38578 & 2.15736 & 1.85279\\
\hline
\end{tabular}
\end{table}

But in the present study, the dissociation temperature is calculate only for the ground states of quarkonia, ${J/\psi}$ and ${\Upsilon}$ at
$\mu=300MeV$ and $N=0$ for the lower bound state of the $T_D$ by applying the condition i.e, $E_{bin}=3(T_D)$ (is in unit of $T_c$).
If the dimensionality number is increased then the values of dissociation temperature and binding energy is also increased. 
The value of anisotropy (\mbox{\boldmath$\xi$}) have been increased in oblate case only then the values of $T_D$ is decreased. 
The values of dissociation temperature given in table-$1$ for QP debye mass and in table-$2$ for LO debye mass is expected for the ground states of quarkonia. Hence we conclude that after noticed the variation of binding energy in figure \ref{fig.7}, we have been conclude that the value of QP debye mass is smaller than that of LO and lattice parameterized form of Debye mass. These observations can be understood from the hierarchy in their numerical values of these debye masses, $m_D^{QP}$ $<$ $m_D^{LO}$ $<$ $m_D^{L}$, and $(B.E)^{QP}$ $<$ $(B.E)^{LO}$ $<$ $(B.E)^{L}$. And this behaviour is also satisfied by the values of table-$1$ and table-$2$ because the dissociation of LO Debye mass is greater then the QP Debye mass. Hence we also conclude that $T_D^{QP}$ $<$ $T_D^{LO}$ $<$ $T_D^{L}$.

\section{Acknowledgments}
One of the authors, VKA acknowledges the Science and Engineering Research Board (SERB) Project No. EEQ/2018/000181 New Delhi for the research support in basic sciences.

\end{document}